\documentclass[12pt]{iopart}

\usepackage{graphicx}
\usepackage{dcolumn}
\usepackage{bm}
\usepackage{psfrag}
\usepackage{subfigure}

\newcommand{\be}{\begin{eqnarray}}
\newcommand{\ee}{\end{eqnarray}}
\newcommand{\bn}{\begin{eqnarray*}}
\newcommand{\en}{\end{eqnarray*}}

\newcommand{\nn}{\nonumber \\}
\newcommand{\nl}{\\}

\renewcommand{\vec}[1]{\mbox{\boldmath$#1$}}
\renewcommand{\d}{\mbox{\rm d}}

\renewcommand{\th}{\ensuremath{\theta}}

\newcommand{\ph}{\ensuremath{\phi}}

\newcommand{\al}{\ensuremath{\alpha}}
\newcommand{\bt}{\ensuremath{\beta}}
\newcommand{\sg}{\ensuremath{\sigma}}
\newcommand{\gm}{\ensuremath{\gamma}}
\newcommand{\dl}{\ensuremath{\delta}}
\newcommand{\lm}{\ensuremath{\lambda}}

\newcommand{\Dl}{\ensuremath{\Delta}}

\newcommand{\Gm}{\ensuremath{\Gamma}}
\newcommand{\Om}{\ensuremath{\Omega}}
\newcommand{\OmK}{\ensuremath{\Omega_{\rm K}}}
\newcommand{\omz}{\ensuremath{\omega}_{\rm zitt.}}

\newcommand{\ze}{\ensuremath{\hat{0}}}

\newcommand{\pvec}{\ensuremath{\vec{p}}}
\newcommand{\Pvec}{\ensuremath{\vec{P}}}

\newcommand{\Rvec}{\ensuremath{\vec{R}}}

\newcommand{\alvec}{\ensuremath{\vec{\al}}}
\newcommand{\sgvec}{\ensuremath{\vec{\sg}}}

\newcommand{\Avec}{\ensuremath{\vec{A}}}

\newcommand{\Xvec}{\ensuremath{\vec{X}}}
\newcommand{\Wvec}{\ensuremath{\vec{W}}}
\newcommand{\Gmvec}{\ensuremath{\vec{\Gm}}}
\newcommand{\Omvec}{\ensuremath{\vec{\Omega}}}

\newcommand{\etavec}{\ensuremath{\vec{\eta}}}
\newcommand{\Pivec}{\ensuremath{\vec{\Pi}}}

\newcommand{\nabvec}{\ensuremath{\vec{\nabla}}}

\newcommand{\hb}{\ensuremath{\hbar}}
\newcommand{\ihb}{\ensuremath{i \hbar}}
\renewcommand{\pt}[1]{\ensuremath{{\partial \over \partial #1}}}

\newcommand{\lt}{\ensuremath{\left}}
\newcommand{\rt}{\ensuremath{\right}}

\renewcommand{\d}{\mbox{\rm d}}

\begin{document}

\title[Effects of Space-Time Curvature on Spin-1/2 Particle {\em Zitterbewegung}]
{Effects of Space-Time Curvature on Spin-1/2 Particle {\em Zitterbewegung}}

\author{Dinesh Singh\dag
\footnote[3]{Author to whom correspondence should be addressed.} \, and
Nader Mobed\dag
}

\address{\dag\ Department of Physics, University of Regina,
Regina, Saskatchewan, S4S 0A2, Canada}

\eads{\mailto{dinesh.singh@uregina.ca} and \mailto{nader.mobed@uregina.ca}
}

\begin{abstract}
This paper investigates the properties of spin-1/2 particle {\em Zitterbewegung} in the presence of
a general curved space-time background described in terms of Fermi normal co-ordinates, where the spatial
part is expressed using general curvilinear co-ordinates.
Adopting the approach first introduced by Barut and Bracken for {\em Zitterbewegung} in the
local rest frame of the particle, it is shown that non-trivial gravitational contributions to the relative position
and momentum operators appear due to the coupling of {\em Zitterbewegung} frequency terms with the Ricci curvature tensor
in the Fermi frame, indicating a formal violation of the weak equivalence principle.
Explicit expressions for these contributions are shown for the case of quasi-circular orbital motion
of a spin-1/2 particle in a Vaidya background.
Formal expressions also appear for the time-derivative of the Pauli-Lubanski vector due to
space-time curvature effects coupled to the {\em Zitterbewegung} frequency.
As well, the choice of curvilinear co-ordinates results in non-inertial contributions in the time evolution of the
canonical momentum for the spin-1/2 particle, where {\em Zitterbewegung} effects
lead to stability considerations for its propagation, based on the Floquet theory of differential equations.
\end{abstract}

\pacs{03.65.Ta, 04.20.-q, 04.20.Cv}






\section{Introduction}
\label{sec:1}

The concept of {\em Zitterbewegung}, or ``trembling motion'' in German due to the rapid
oscillation of a spin-1/2 particle about its classical worldline, is one that still generates
great interest about foundational issues in quantum mechanics, even after over 70 years since
its initial discovery by Schr\"{o}dinger \cite{Schrodinger}.
When investigating the free-particle motion of an electron via the Dirac equation, he showed that
its instantaneous velocity $v_{\rm inst.}$ equals the speed of light $c$, which clearly
disagrees with the fact that a massive particle's observed speed will always be $v_{\rm obs.} < c$.
However, Dirac convincingly emphasized \cite{Dirac} that this apparent contradiction is resolved by
noting that $v_{\rm obs.}$ is really an {\em average velocity} determined by measuring the electron's
position over a sufficiently small time interval $\Dl t$.
When $c \Dl t \gg \hb/mc$, the Compton wavelength for the electron of mass $m$, it follows that
the practical measurement of $v_{\rm obs.}$ always leads to a preservation of causality, and without
any complications due to {\em Zitterbewegung}.
Nevertheless, this highly non-intuitive oscillatory behaviour identified to exist for massive
spin-1/2 particles is a conceptually rich feature of relativistic quantum mechanics that still offers
both thought-provoking challenges and opportunities for deeper exploration \cite{Sidharth,Wang,Hestenes},
especially when applied to more complicated situations.

While past and recent investigations of {\em Zitterbewegung} since Schr\"{o}dinger's and Dirac's studies
have been largely motivated by attempts to understand the intrinsic structure of the electron \cite{Barut1},
there currently exists an intensive practical study of the concept within the context
of condensed-matter theory and experiment.
Recent explorations of this phenomenon involving two-dimensional carbon sheets known as {\em graphene}
have been proposed \cite{Katsnelson,Cserti}, where {\em Zitterbewegung} effects are potentially
observable in the non-relativistic limit of the Dirac equation, while a second paper claims
a direct observation for photons in a two-dimensional photonic crystal \cite{Zhang}, and a third
paper proposes detection of {\em Zitterbewegung} using ultracold neutral atoms \cite{Vaishnav}.
These studies are especially interesting given the largely accepted understanding that free-particle
{\em Zitterbewegung} arises only in situations where positive- and negative-energy solutions to the Dirac equation
can interact, allowing for quantum interference effects to manifest the rapid oscillation terms
that would otherwise remain decoupled and unobservable.

This leads to the following consideration where it concerns interactions in a gravitational background.
Given that external fields can introduce non-trivial effects that lead to observable consequences
for electron {\em Zitterbewegung}, and given that gravitation is best described (for now) by Einstein's theory
of general relativity in terms of the {\em dynamical} curvature of space-time due to external matter,
it is reasonable to envision curved space-time as something like a {\em medium} with an energy density content
not unlike that of an electromagnetic background.
Of course, there is the fundamental distinction that Einstein's {\em equivalence principle} is an integral part of
general relativity, reflected mathematically as diffeomorphism invariance under co-ordinate transformations.
This is certainly reasonable when considering purely {\em classical} descriptions of matter,
where it is meaningful to speak about its localization within a compact spatial region.
However, conceptual difficulties arise when considering {\em quantum matter} in a curved space-time
background, where issues of non-locality due to the wave-particle duality of quantum objects overlap
with {\em classical} descriptions of space-time curvature.
This is a particularly relevant issue considering that gravitational effects are manifested locally
from observing tidal forces via the geodesic deviation equation, which requires the existence of
a {\em neighbouring} geodesic in relation to the reference worldline.
It is very unclear how to disentangle the competing effects of quantum non-locality
and the locality of classical gravitation in a way that results in meaningful statements about
quantum matter effects in a curved space-time background, including {\em Zitterbewegung}.

The purpose of this paper is to explore the physical properties of spin-1/2 particle
{\em Zitterbewegung} in a curved space-time background described locally in terms of Fermi normal
co-ordinates.
Its motivation is based on understanding whether spatial fluctuations about the particle's
worldline lead to non-trivial gravitational corrections that influence its propagation in space-time.
A recent innovation in the standard formalism of Fermi normal co-ordinates
is to describe the spatial fluctuations in terms of curvilinear spatial co-ordinates \cite{Singh-Mobed1,Singh-Mobed2,Singh1}.
This paper begins with a brief review of the existing treatment of {\em Zitterbewegung}
in a flat space-time background, as given in Section~\ref{sec:2}.
Because the Fermi normal co-ordinate system is defined locally about the particle's worldline,
it is important to make use of a local perspective for the occurrence of {\em Zitterbewegung}
and explore its physical consequences.
It so happens that the approach introduced by Barut and Bracken \cite{Barut2}
is particularly well-suited for this purpose, and so this perspective is adopted
for consideration in a curved space-time background.
This leads to a brief introduction of the covariant Dirac equation in Fermi normal co-ordinates
in Section~\ref{sec:3}, which includes the curvilinear co-ordinate generalization.
A presentation of the corresponding Dirac Hamiltonian derived from within this framework
is given in Section~\ref{sec:4}, where contributions due to {\em Zitterbewegung} reveal
a formal violation of the weak equivalence principle due to direct coupling of the gravitational
field with the particle's mass.
From this Hamiltonian, it becomes possible to compute the time-evolution of quantum operators
in Section~\ref{sec:5} and show how {\em Zitterbewegung} formally appears in the expressions.
Particular attention is given to the time-evolution of the position and momentum operators,
and also the covariant spin operator described by the Pauli-Lubanski vector.
A brief conclusion follows in Section~\ref{sec:6}.

For this paper, geometric units of $G = c = 1$ are adopted, while the curvature tensor definitions
follow the conventions given by Misner, Thorne, and Wheeler \cite{MTW}, but with metric signature $-2$.
As well, the flat space-time gamma matrices follow the conventions of Itzykson and Zuber \cite{Itzykson}.

\section{Review of Free-Particle Spin-1/2 {\em Zitterbewegung} in Flat Space-Time}
\label{sec:2}

\subsection{General Formalism}

Before introducing {\em Zitterbewegung} in curved space-time, it is useful to first review
the original treatment of the problem in flat space-time by Schr\"{o}dinger \cite{Schrodinger,Dirac,Barut2}.
Assuming space-time co-ordinates in Cartesian form $X^{\hat{\mu}} = \lt(T, \Xvec\rt)$,
the Schr\"{o}dinger equation is $i \hbar \lt(\partial/\partial T \rt) \psi(X) = H_0 \, \psi(X)$,
where
\be
H_0 & = & m \bt_{(0)} + \alvec_{(0)} \cdot \Pvec_{(0)} \,
\label{H0}
\ee
is the free-particle Dirac Hamiltonian defined in terms of spin-1/2 particle mass $m$, canonical momentum
$\Pvec^{(0)}_{\hat{\mu}} = i \hb \, \partial/\partial X^{\hat{\mu}}$, and
$4 \times 4$ matrices $\alvec_{(0)}^{\hat{\jmath}} = \gm^{\hat{0}} \, \gm^{\hat{\jmath}}$
and $\bt_{(0)} = \gm^{\hat{0}}$ with $\lt\{\gm^{\hat{\al}} \, , \gm^{\hat{\bt}}\rt\} = 2 \eta^{{\hat{\al}} {\hat{\bt}}}$ \cite{Itzykson}.
As expected, the canonical momentum operators satisfy
\be
\lt[\Pvec^{(0)}_{\hat{\mu}} \, , \Pvec^{(0)}_{\hat{\nu}}\rt] & = & 0 \, ,
\label{P-commutator}
\ee
while the corresponding position operators $\Xvec_{(0)}^{\hat{\jmath}}$ satisfy
\be
\lt[\Xvec_{(0)}^{\hat{\imath}} \, , \Xvec_{(0)}^{\hat{\jmath}}\rt] & = & 0 \, ,
\qquad \lt[\Xvec_{(0)}^{\hat{\imath}} \, , \Pvec^{(0)}_{\hat{\jmath}}\rt] \ = \ -i \hb \, \dl^{\hat{\imath}}{}_{\hat{\jmath}}
\, .
\label{X-commutator}
\ee

Adopting the Heisenberg picture, it follows that the time derivative of some operator $\Avec$ is
\be
{\d \Avec(T) \over \d T} & = & {i \over \hb} \lt[H_0 \, , \Avec(T)\rt] \, .
\label{dA-dt}
\ee
Clearly, $\d \Pvec_{(0)}^{\hat{\jmath}}(T)/\d T = \d H_0/\d T = 0$, while the instantaneous velocity and acceleration operators are
\be
{\d \Xvec_{(0)}^{\hat{\jmath}}(T) \over \d T} & = & \alvec_{(0)}^{\hat{\jmath}}(T) \, ,
\label{dx-dt}
\nl
\nn
{\d^2 \Xvec_{(0)}^{\hat{\jmath}}(T) \over \d T^2} & = & {\d \alvec_{(0)}^{\hat{\jmath}}(T) \over \d T} \ = \
-{2i \over \hb} \, \etavec_{(0)}^{\hat{\jmath}}(T) \, H_0 \, ,
\label{dv-dt}
\ee
where
\be
\etavec_{(0)}^{\hat{\jmath}}(T) & = & \alvec_{(0)}^{\hat{\jmath}}(T) - \Pvec_{(0)}^{\hat{\jmath}}(T) \, H_0^{-1} \, .
\label{eta}
\ee
Because $\Pvec_{(0)}^{\hat{\jmath}}(T)$ and $H_0$ are constants of the motion, it is also true that
\be
{\d \etavec_{(0)}^{\hat{\jmath}}(T) \over \d T} & = & {\d \over \d T}
\lt[\alvec_{(0)}^{\hat{\jmath}}(T) - \Pvec_{(0)}^{\hat{\jmath}}(T) \, H_0^{-1}\rt]
 \ = \ {\d \alvec_{(0)}^{\hat{\jmath}}(T) \over \d T}
\nn
& = &
-{2i \over \hb} \, \etavec_{(0)}^{\hat{\jmath}}(T) \, H_0 \, .
\label{d-eta/dt}
\ee
An immediate solution follows for $\etavec_{(0)}^{\hat{\jmath}}(T)$, such that
\be
\etavec_{(0)}^{\hat{\jmath}}(T) & = & \etavec_{(0)}^{\hat{\jmath}}(0) \, e^{-2i H_0 T/\hb} \, ,
\label{eta0-solution}
\ee
or
\be
\alvec_{(0)}^{\hat{\jmath}}(T) & = & \Pvec_{(0)}^{\hat{\jmath}}(0) \, H_0^{-1} + \etavec_{(0)}^{\hat{\jmath}}(0) \, e^{-2i H_0 T/\hb} \, .
\label{alpha-free-particle}
\ee
Direct integration of (\ref{alpha-free-particle}) then leads to the free-particle solution for $\Xvec_{(0)}^{\hat{\jmath}}(T)$, such that
the time-evolution for the position and canonical momentum operators are
\be
\Xvec_{(0)}^{\hat{\jmath}}(T) & = & \Xvec_{(0)}^{\hat{\jmath}}(0) + \lt(\Pvec_{(0)}^{\hat{\jmath}}(0) \, H_0^{-1}\rt) T
\nn
&  &{} + {\hb \over 2} \, \etavec_{(0)}^{\hat{\jmath}}(0) \, H_0^{-1} \lt[\sin \lt(2 H_0 T/\hb\rt) - 2i \, \sin^2\lt(H_0 T/\hb\rt)\rt] \, ,
\label{X-free-particle}
\nl
\nn
\Pvec_{(0)}^{\hat{\jmath}}(T) & = & \Pvec_{(0)}^{\hat{\jmath}}(0) \, .
\label{P-free-particle}
\ee
The first two terms of (\ref{X-free-particle}) denote the expected mean position as a function of time $T$,
while the remaining terms show the rapid oscillation effect superimposed about the spin-1/2 particle's worldline
due to {\em Zitterbewegung}.
For future reference, it is straightforward to show that {\em Zitterbewegung} also applies to
$\bt_{(0)}(T)$ and the Pauli spin matrix $\sgvec_{(0)}^{\hat{\jmath}}(T)$, such that
\be
\fl
\bt_{(0)}(T) & = & \bt_0 \, e^{-2i H_0 T/\hbar}
+ m \, H_0^{-1} \lt[2 \, \sin^2 \lt(H_0 T/\hb\rt) + i \, \sin\lt(2 H_0 T/\hb\rt)\rt] \, ,
\label{beta-free-particle}
\nl
\nn
\fl
\sgvec_{(0)}^{\hat{\jmath}}(T) & = & \sgvec_0^{\hat{\jmath}}
+ i \lt(\alvec_0 \times \Pvec_{(0)}(0)\rt)^{\hat{\jmath}} \, H_0^{-1}
\lt[2 \, \sin^2 \lt(H_0 T/\hb\rt) + i \, \sin\lt(2 H_0 T/\hb\rt)\rt] \, ,
\label{sigma-free-particle}
\ee
where
\be
\alvec_0^{\hat{\jmath}} & \equiv & \alvec_{(0)}^{\hat{\jmath}}(0) \, , \qquad \bt_0 \ \equiv \ \bt_{(0)}(0) \, , \qquad
\sgvec_0^{\hat{\jmath}} \ \equiv \ \sgvec_{(0)}^{\hat{\jmath}}(0) \, .
\label{alpha0-sigma0}
\ee


Before continuing, it is worthwhile to explore the mathematical details of the operators which describe the
{\em Zitterbewegung} phenomenon.
Based on a precisely formulated description of the problem \cite{Thaller}, it is possible to establish
that the domain ${\cal D}$ corresponding to the position operator $\Xvec_{(0)}^{\hat{\jmath}}(0)$
is invariant under time evolution, such that
\be
{\cal D} \lt(\Xvec_{(0)}^{\hat{\jmath}}(T)\rt) & = & e^{-i H_0 T/\hbar} \, {\cal D} \lt(\Xvec_{(0)}^{\hat{\jmath}}(0)\rt)
\ = \ {\cal D} \lt(\Xvec_{(0)}^{\hat{\jmath}}(0)\rt) \, .
\label{domain-X}
\ee
As such, (\ref{domain-X}) implies that $\Xvec_{(0)}^{\hat{\jmath}}(T)$ remains bounded for all finite $T$.
In addition, the expectation value for the time-evolved position operator is well behaved, in that
the component of (\ref{X-free-particle}) exhibiting {\em Zitterbewegung} integrates to zero in the expectation value
as $T \rightarrow \infty$.
Furthermore, it is well-known \cite{Thaller,Sakurai,Greiner} that the Hilbert space identified with solutions
of the free-particle Dirac equation decouples into invariant subspaces defined by positive and negative energy states,
respectively.
This implies that the expectation values of $\Xvec_{(0)}^{\hat{\jmath}}(T)$
formed by exclusively positive energy or negative energy states never reveal any {\em Zitterbewegung}-induced effects.
Not only does it explain why {\em Zitterbewegung} is not an observable for a propagating free particle,
it suggests the existence of a breakdown in the single-particle treatment of quantum matter, justifying
the need for the standard quantum field theory description.

\subsection{The Barut-Bracken Approach to {\em Zitterbewegung}}

The main expressions above can apply to spin-1/2 particle motion in a general reference frame with $\Pvec_{(0)}^{\hat{\jmath}}(T) \neq 0$.
Schr\"{o}dinger's original view of {\em Zitterbewegung} was to search for ``microscopic'' degrees of freedom
for position, momentum, and angular momentum to explain the electron's rapid oscillatory behaviour as a separable effect
superimposed on its observable motion \cite{Schrodinger,Barut2}.
To this end, he proposed the existence of a {\em microscopic position} operator $\vec{\xi}^{\hat{\jmath}}(T)$, according to
\be
\Xvec_{(0)}^{\hat{\jmath}}(T) & = & \Xvec_{(A)}^{\hat{\jmath}}(T) + \vec{\xi}^{\hat{\jmath}}(T) \, ,
\label{X-Schrodinger}
\ee
where
\be
\Xvec_{(A)}^{\hat{\jmath}}(T) & = & \Xvec_{(0)}^{\hat{\jmath}}(0) + \lt(\Pvec_{(0)}^{\hat{\jmath}}(0) \, H_0^{-1}\rt) T
- {i \hb \over 2} \, \etavec_{(0)}^{\hat{\jmath}}(0) \, H_0^{-1} \, ,
\label{X-A}
\nl
\vec{\xi}^{\hat{\jmath}}(T) & = & {i \hb \over 2} \, \etavec_{(0)}^{\hat{\jmath}}(0) \, H_0^{-1} \,
e^{-2i H_0 T/\hb} \, ,
\label{xi}
\ee
along with a {\em microscopic momentum} associated with $\vec{\xi}^{\hat{\jmath}}(T)$, in the form
\be
\etavec_{(0)}^{\hat{\jmath}}(T) \, H_0 + \Pvec_{(0)}^{\hat{\jmath}}(T) & = &
{\d \Xvec_{(0)}^{\hat{\jmath}}(T) \over \d T} \, H_0 \, .
\label{P-Schrodinger-1}
\ee

In re-examining {\em Zitterbewegung} with the motive to understand the electron's intrinsic structure
for addressing self-energy and renormalization issues in QED, Barut and Bracken \cite{Barut2} saw no difficulty with
Schr\"{o}dinger's formulation of microscopic position, according to (\ref{X-Schrodinger}).
However, they had serious issues with his definition of microscopic momentum (\ref{P-Schrodinger-1}),
primarily motivated by the claim that an equally attractive--but incompatible--definition could be proposed
within his formalism, such that
\be
H_0 \, \etavec_{(0)}^{\hat{\jmath}}(T) + \Pvec_{(0)}^{\hat{\jmath}}(T) & = &
H_0 \, {\d \Xvec_{(0)}^{\hat{\jmath}}(T) \over \d T} \, .
\label{P-Schrodinger-2}
\ee
Another major deficiency is that (\ref{P-Schrodinger-1}) and (\ref{P-Schrodinger-2}) are {\em not} Hermitian operators,
and that it is impossible to construct a viable Hermitian combination of these two momentum definitions.
As well, because the canonical momentum $\Pvec_{(0)}^{\hat{\jmath}}(T)$ is a constant of the motion,
they argue that it should represent the total overall momentum of the system, and that the {\em microscopic}
nature of (\ref{P-Schrodinger-1}) should not have any dependence on an inherently
{\em macroscopic} description of momentum.

To deal with these conceptual deficiencies identified by Barut and Bracken in Schr\"{o}dinger's
treatment of {\em Zitterbewegung}, they proposed the existence
of a {\em relative momentum} operator $\vec{\cal P}_{(B)}^{\hat{\jmath}}(T)$ that results from performing the computations
in the local rest frame.\footnote{The main advantage of Barut's and Bracken's approach over Schr\"{o}dinger's is that the derived
relative position and momentum operators form a harmonic oscillator relationship with frequency $\omz$
and are part of a basis set satisfying an SO(5) Lie algebra structure \cite{Barut2}.}
This amounts to letting the operators $\Avec_{(0)}(T) \rightarrow \Avec_{(B)}(T)$, where
\be
\Pvec_{(B)}^{\hat{\jmath}}(T) & = & 0 \, .
\label{P0=0-Barut}
\ee
By also setting the initial position to $\Xvec_{(B)}^{\hat{\jmath}}(0) = 0$, a {\em relative position} operator
in the local frame can be identified, such that
\be
\Xvec_{(B)}^{\hat{\jmath}}(T) & \approx & {\hb \over 2m} \lt[\sin\lt(\omz T\rt) + 2i \, \sin^2\lt(\omz T/2\rt) \bt_0\rt] \alvec_0^{\hat{\jmath}} \, ,
\label{rel-X-Barut}
\ee
while the remaining operators (\ref{alpha-free-particle}), (\ref{beta-free-particle}), and (\ref{sigma-free-particle}) become
\be
\alvec_{(B)}^{\hat{\jmath}}(T) & \approx & \alvec_0^{\hat{\jmath}} \lt[\cos\lt(\omz T\rt) - i \, \sin\lt(\omz T\rt) \, \bt_0\rt] \, ,
\label{alpha-Barut}
\nl
\bt_{(B)}(T) & \approx & \bt_0 \, ,
\label{beta-Barut}
\nl
\sgvec_{(B)}^{\hat{\jmath}}(T) & \approx & \sgvec_0^{\hat{\jmath}} \, ,
\label{sigma-Barut}
\ee
where $\omz = 2m/\hb$ is the {\em Zitterbewegung} frequency derived from
\be
H_0 & = & m \bt_{(B)}(T) \ \approx \ m \, \bt_0 \, ,
\label{H0-Barut}
\nl
H_0^{-1} & \approx & m^{-1} \, \bt_0 \, .
\label{H0-inv-Barut}
\ee
Barut and Bracken subsequently propose that the relative momentum is described by
\be
\vec{\cal P}_{(B)}^{\hat{\jmath}}(T) & = & m \, \alvec_{(B)}^{\hat{\jmath}}(T) \,
\label{rel-P-Barut}
\ee
with respect to the spin-1/2 particle's classical worldline.

It should be noted that $\Xvec_{(B)}^{\hat{\jmath}}(T)$ differs slightly from $\vec{\xi}^{\hat{\jmath}}(T)$
incorporated in the local rest frame by Barut and Bracken \cite{Barut2}.
This is due to a minor change in definition, such that the third term of (\ref{X-A})
is absorbed by (\ref{xi}) to form (\ref{rel-X-Barut}), instead of expressing (\ref{X-free-particle})
as given by (\ref{X-Schrodinger})--(\ref{xi}) in the local rest frame, where
\be
\fl
\Xvec_{(A)}^{\hat{\jmath}}(T) & = & {i \hb \over 2m} \, \bt_0 \, \alvec_0^{\hat{\jmath}} \, ,
\label{X-A-rest-frame}
\nl
\nn
\fl
\vec{\xi}^{\hat{\jmath}}(T) & = & -{i \hb \over 2m} \, \bt_0 \, \alvec_0^{\hat{\jmath}} \, e^{-i \, \omz \, T \, \bt_0}
\ = \ {\hb \over 2m}
\lt[\sin\lt(\omz T\rt) - i \cos\lt(\omz T\rt) \bt_0 \rt] \alvec_0^{\hat{\jmath}}
\nn
& = & {\hb \over 2m} \, e^{i \, \omz \lt(T - \pi/2\rt) \bt_0} \, \alvec_0^{\hat{\jmath}} \, .
\label{xi-rest-frame}
\ee

With the definitions for relative position and momentum now given, it is possible to derive the uncertainty in their measurements,
to compare with the Heisenberg uncertainty relation.
To begin, consider the time-averaged operator over a complete cycle in terms of $\omz$, such that
\be
\lt\langle \Avec(T) \rt\rangle & \equiv & {\omz \over 2 \pi} \, \int_0^{2 \pi/\omz} \Avec(T) \, \d T \, .
\label{zitt-time-avg-A}
\ee
It follows that the {\em Zitterbewegung} time average of (\ref{rel-X-Barut}) is
\be
\lt\langle \Xvec_{(B)}^{\hat{\jmath}}(T) \rt\rangle & = & {i \hb \over 2m} \, \bt_0 \, \alvec_0^{\hat{\jmath}} \, ,
\label{X-Barut-magnitude-component}
\ee
and subsequently leads to
\be
\lt|\lt\langle \Xvec_{(B)}(T) \rt\rangle\rt|  & = &
\lt[-\eta_{{\hat{\imath}}{\hat{\jmath}}} \, \lt\langle \Xvec_{(B)}^{\hat{\imath}}(T) \rt\rangle \lt\langle \Xvec_{(B)}^{\hat{\jmath}}(T) \rt\rangle\rt]^{1/2}
\ = \ {\sqrt{3} \over 2} \lt(\hb \over m\rt) \, ,
\label{X-Barut-magnitude-time-avg}
\ee
in agreement with an earlier computation derived in a similar context \cite{Barut3}.
However, the squared-magnitude of $\Xvec_{(B)}^{\hat{\jmath}}(T)$ prior to time averaging is
\be
\lt|\Xvec_{(B)}(T)\rt|^2 & = & - \eta_{{\hat{\imath}}{\hat{\jmath}}} \, \Xvec_{(B)}^{\hat{\imath}}(T) \, \Xvec_{(B)}^{\hat{\jmath}}(T)
\ = \ {12 \over \omz^2} \, \sin^2 \lt(\omz T/2\rt) \, ,
\label{X-Barut-magnitude-sq}
\ee
which leads to
\be
\lt\langle \lt|\Xvec_{(B)}(T)\rt|^2 \rt\rangle^{1/2} & = & \sqrt{3 \over 2} \lt(\hb \over m\rt) \, .
\label{X-Barut-magnitude-sq-time-avg-root}
\ee
Therefore, the local uncertainty in relative position is
\be
\Dl \Xvec_{(B)} & = & \sqrt{\lt\langle \lt|\Xvec_{(B)}(T)\rt|^2 \rt\rangle - \lt|\lt\langle \Xvec_{(B)}(T) \rt\rangle\rt|^2}
\ = \ {\sqrt{3} \over 2} \lt(\hb \over m\rt) \, .
\label{dX-Barut}
\ee

To compute the corresponding local uncertainty in the relative momentum, it is self-evident that
\be
\lt\langle \vec{\cal P}_{(B)}^{\hat{\jmath}}(T) \rt\rangle & = & 0 \, ,
\label{rel-P-Barut-magnitude-component}
\ee
while
\be
\lt\langle \lt|\vec{\cal P}_{(B)}(T)\rt|^2 \rt\rangle^{1/2} & = & \sqrt{3} \, m \, .
\label{rel-P-Barut-magnitude-sq-time-avg-root}
\ee
This results in $\Dl \vec{\cal P}_{(B)} = \sqrt{3} \, m$,
leading to the expression
\be
\lt(\Dl \Xvec_{(B)}\rt) \lt(\Dl \vec{\cal P}_{(B)}\rt) & = & 3 \lt(\hb \over 2\rt) \, ,
\label{Heisenberg-Barut}
\ee
well in agreement with the Heisenberg position-momentum uncertainty relation
$\lt(\Dl \Xvec_{(B)}\rt) \lt(\Dl \vec{\cal P}_{(B)}\rt) \geq \hb/2$.

\section{An Approach to the Curved Space-Time Generalization}
\label{sec:3}

\subsection{The Hypothesis of Locality}

Fundamental to the requirements of this paper is consideration of the implicitly accepted {\it hypothesis of locality} \cite{Mashhoon}
for classical phenomena and how it relates to the intersection of quantum mechanical behaviour in curved space-time.
This hypothesis essentially states that, at any {\em instantaneous moment} of proper time on a classical worldline,
it is possible to identify an {\em instantaneous comoving inertial frame} tangent to the worldline.
A smooth one-to-one correspondence follows from this identification, along with an intrinsic length and time scale
correlated with the comoving observer.

When considering purely {\em classical matter}, this hypothesis is undeniably successful.
Issues become much more ambiguous, however, when applying this hypothesis to single-particle quantum states--let alone
quantum fields, since the wave-particle duality implies the need for a {\em region} of space-time to establish
the location of quantum matter that is consistent with Heisenberg's uncertainty principle.
This especially applies to quantum matter with long wavelengths when compared to the characteristic length
of the background gravitational source, where quantum interference effects become most relevant to disentangle conceptually.
If a measurement applied to quantum matter is somehow performed on a sufficiently small time scale compared to
the comoving observer's intrinsic time scale, it is reasonable to surmise that the hypothesis of locality still applies.
However, this claim is still a tenuous one, at best, and should be subjected to deeper analysis.

\subsection{Covariant Dirac Equation in Fermi Normal Co-ordinates}

Accepting the hypothesis of locality as a viable starting point, the approach to {\em Zitterbewegung}
in curved space-time is to adopt Fermi normal co-ordinates $X^\mu$ with respect to a local comoving frame \cite{Singh-Mobed1},
with $X^0 = T$ identified as the spin-1/2 particle's proper time and $X^j$ as the Cartesian spatial co-ordinates
orthogonal to the worldline.
Again, for purely classical phenomena, this choice for $X^j$ is perfectly reasonable.
However, when considering the propagation of quantum matter, it may be more appropriate to use
general curvilinear co-ordinates $U^\mu = \lt(T, u^j\rt)$ to better reflect the symmetries associated with
the particle's trajectory in space for general motion between neighbouring intervals of proper time,
where $X^j = X^j(u)$ \cite{Singh-Mobed1}.
Eventually, the spatial Fermi normal co-ordinates will be identified with {\em quantum fluctuations}
about the classical worldline, such that $X^j \rightarrow \Xvec_{(B)}^{\hat{\jmath}}(T)$ when
deriving the Dirac Hamiltonian.
This is the central hypothesis that results in the description of curved space-time {\em Zitterbewegung} to follow.

Starting with a worldline ${\cal C}$ defined in a general space-time background and parametrized by proper time $\tau$,
the Fermi frame \cite{Manasse,Synge,Mashhoon1,Mashhoon2} is determined at some event $P_0$ on ${\cal C}$ by
constructing a local orthonormal vierbein set $\lt\{ \lm^{\bar{\alpha}}{}_\mu \rt\} \,$ and inverse set $\lt\{ \lm^\mu{}_{\bar{\alpha}} \rt\}$.
Then the local spatial axes are defined by $\lm^{\bar{\mu}}{}_0 = \d x^{\bar{\mu}}/\d \tau$ and $\lm^{\bar{\mu}}{}_a$.
If $\xi^{\bar{\mu}} = \lt(\d x^{\bar{\mu}}/\d \sg\rt)_0$ denotes the unit spatial tangent vector from $P_0$ to a neighbouring event $P$,
where a unique spacelike geodesic orthogonal to ${\cal C}$ exists along proper length $\sg$,
then $T = \tau$ and $X^i = \sg \, \xi^{\bar{\mu}} \, \lm^i{}_{\bar{\mu}}$ become the Fermi normal co-ordinates at $P$.
The corresponding space-time metric is then described by
\be
\d s^2 & = & {}^F{}g_{\mu \nu}(X) \, \d X^\mu \, \d X^\nu \, ,
\label{F-metric}
\ee
where
\numparts
\be
{}^F{}g_{00}(X) & = & 1 + {}^F{}R_{0i0j}(T) \, X^i \, X^j + \cdots \, ,
\label{F-g00}
\nl
{}^F{}g_{0j}(X) & = & {2 \over 3} \, {}^F{}R_{0ijk}(T) \, X^i \, X^k + \cdots \, ,
\label{F-g0j}
\nl
{}^F{}g_{ij}(X) & = & \eta_{ij} + {1 \over 3} \, {}^F{}R_{ikjl}(T) \, X^k \, X^l + \cdots \, ,
\label{F-gij}
\ee
\endnumparts
and
\be
{}^F{}R_{\al \bt \gm \dl}(T) & = & R_{\bar{\mu} \bar{\nu} \bar{\rho} \bar{\sg}} \,
\lm^{\bar{\mu}}{}_\al \, \lm^{\bar{\nu}}{}_\bt \, \lm^{\bar{\rho}}{}_{\gm} \, \lm^{\bar{\sg}}{}_\dl \, .
\label{F-Riemann}
\ee

The covariant Dirac equation for a spin-1/2 particle with mass $m$ can be written in terms of (\ref{F-g00})--(\ref{F-gij}) as
\be
\lt[i \gm^\mu(X) \lt(\partial_\mu + i \, \Gamma_\mu(X) \rt) - m/\hb\rt]\psi(X) & = & 0 \, ,
\label{Dirac-eq}
\ee
where $\partial_\mu = \partial/\partial X^\mu$
and $\Gamma_\mu(X)$ is the spin connection defined with respect to $\lt\{ \gm^\mu(X) \rt\}$, the set of gamma matrices satisfying
$\lt\{ \gm^\mu(X), \gm^\nu(X) \rt\} = 2 \, g_F^{\mu \nu}(X)$.
A local Lorentz frame \cite{Poisson} can be determined according to
${}^F{}g_{\mu \nu}(X) = \eta_{\hat{\alpha}\hat{\beta}} \, \bar{e}^{\hat{\alpha}}{}_\mu (X) \, \bar{e}^{\hat{\beta}}{}_\nu (X)$,
where $\lt\{ \bar{e}^\mu{}_{\hat{\alpha}}(X) \rt\}$ and $\lt\{ \bar{e}^{\hat{\alpha}}{}_\mu (X)\rt\} \,$
form a respective orthonormal vierbein and inverse vierbein set, such that
\numparts
\be
\bar{e}^{\hat{0}}{}_0(X) & = & 1 + {1 \over 2} \, {}^F{}R_{0i0j}(T) \, X^i \, X^j \, ,
\label{inv-tetrad-00}
\nl
\bar{e}^{\hat{0}}{}_j(X) & = & {1 \over 6} \, {}^F{}R_{0ijk}(T) \, X^i \, X^k \, ,
\label{inv-tetrad-0j}
\nl
\bar{e}^{\hat{\imath}}{}_0(X) & = & -{1 \over 2} \, {}^F{}R^i{}_{j0k}(T) \, X^j \, X^k \, ,
\label{inv-tetrad-i0}
\nl
\bar{e}^{\hat{\imath}}{}_j(X) & = & \dl^i{}_j - {1 \over 6} \, {}^F{}R^i{}_{kjl}(T) \, X^k \, X^l \, ,
\label{inv-tetrad-ij}
\ee
\endnumparts
and
\numparts
\be
\bar{e}^0{}_{\hat{0}}(X) & = & 1 - {1 \over 2} \, {}^F{}R_{0i0j}(T) \, X^i \, X^j \, ,
\label{tetrad-00}
\nl
\bar{e}^i{}_{\hat{0}}(X) & = & {1 \over 2} \, {}^F{}R^i{}_{j0k}(T) \, X^j \, X^k \, ,
\label{tetrad-i0}
\nl
\bar{e}^0{}_{\hat{\jmath}}(X) & = & -{1 \over 6} \, {}^F{}R_{0ijk}(T) \, X^i \, X^k \, ,
\label{tetrad-0j}
\nl
\bar{e}^i{}_{\hat{\jmath}}(X) & = & \dl^i{}_j + {1 \over 6} \, {}^F{}R^i{}_{kjl}(T) \, X^k \, X^l \, .
\label{tetrad-ij}
\ee
\endnumparts
The spin connection is then determined to be
\be
\Gm_\mu(X) & = & {i \over 4} \, \gm^\al(X) \lt(\nabla_\mu \gm_\al(X) \rt)
\ = \ -{1 \over 4} \, \sg^{\hat{\al} \hat{\bt}} \, \eta_{\hat{\bt} \hat{\gm}} \, \bar{e}^\al{}_{\hat{\al}} \lt(\nabla_\mu \, \bar{e}^{\hat{\gm}}{}_\al \rt),
\label{spin-connection}
\ee
where $\nabla_\mu$ is the covariant derivative operator and $\sg^{\hat{\al} \hat{\bt}} = {i \over 2} \lt[\gm^{\hat{\al}}, \gm^{\hat{\bt}}\rt]$.
To first-order in the Riemann tensor, it is shown from (\ref{spin-connection}) that
\be
\fl
\Gm_0(X) & = & i \, \gm^{\hat{0}} \, \gm^{\hat{\imath}} \lt[{1 \over 2} \, {}^F{}R_{i00k}(T) + {1 \over 3} \, {}^F{}R_{ij0k,0}(T) \, X^j \rt] X^k \, ,
\label{Gm-0}
\nl
\fl
\Gm_l(X) & = & i \, \gm^{\hat{0}} \, \gm^{\hat{\imath}} \lt[{1 \over 3} \,
\lt[{}^F{}R_{il0k}(T) + {}^F{}R_{i[k0]l}(T)\rt] - {1 \over 12} \, {}^F{}R_{ijkl,0}(T) \, X^j \rt] X^k \, ,
\label{Gm-l}
\ee
where ${}^F{}R_{i[k0]l}(T) = {1 \over 2} \lt[{}^F{}R_{ik0l}(T) - {}^F{}R_{i0kl}(T)\rt]$
denotes antisymmetrization of the middle two indices.

\subsection{Conversion to Curvilinear Co-ordinates}
It is straightforward to introduce a conversion of Fermi normal co-ordinates from locally Cartesian to
general curvilinear co-ordinates, in terms of a new set of orthonormal vierbeins described by
\numparts
\be
e^\bt{}_{\hat{\al}}(U) & = & {\partial U^\bt \over \partial X^\al} \, \bar{e}^\al{}_{\hat{\al}}(X) \, , \qquad
\label{tetrad-coord-trans}
\nl
e^{\hat{\al}}{}_\bt(U) & = & {\partial X^\al \over \partial U^\bt} \, \bar{e}^{\hat{\al}}{}_\al(X) \, .
\label{inv-tetrad-coord-trans}
\ee
\endnumparts
The corresponding spin connection in curvilinear co-ordinates becomes
\be
\Gm_\mu(U) & = & {i \over 4} \, \gm^\al(U) \lt[\nabla_\mu \gm_\al(U) \rt]
\ = \ -{1 \over 4} \, \sg^{\hat{\al} \hat{\bt}} \, \eta_{\hat{\bt} \hat{\gm}} \, e^\al{}_{\hat{\al}} \lt(\nabla_\mu \, e^{\hat{\gm}}{}_\al \rt)
\nn
& = & {\partial X^\al \over \partial U^\mu} \, \Gm_\al(X) \, .
\label{spin-connection-curvilinear}
\ee
It follows that the covariant Dirac equation (\ref{Dirac-eq}), expressed in curvilinear co-ordinates
and projected onto the local Lorentz frame, is
\be
\lt[i \gm^{\hat{\mu}} \lt(\hat{\nabla}_{\hat{\mu}} + i \, \Gm_{\hat{\mu}}(U) \rt) - m/\hb\rt]\psi(U) & = & 0 \, ,
\label{Dirac-eq-curvilinear}
\ee
where $\hat{\nabla}_{\hat{\mu}} = \nabvec_{\hat{\mu}} + i \, \hat{\Gm}_{\hat{\mu}}(U)$ is the flat space-time
covariant derivative operator in curvilinear co-ordinate form.
The corresponding line element is represented by
\be
\d s^2 & = & \d T^2 + \eta_{\hat{\imath}\hat{\jmath}} \lt(\lm^{(\hat{\imath})}(u) \, \d u^{\hat{\imath}}\rt)
\lt(\lm^{(\hat{\jmath})}(u) \, \d u^{\hat{\jmath}}\rt) \, ,
\label{Minkowski-curvilinear}
\ee
where $\lm^{(\hat{\imath})}(u)$ are dimensional scale functions \cite{Hassani}, and
\be
\nabvec_{\hat{0}} & \equiv & {\partial \over \partial T} \, , \qquad
\nabvec_{\hat{\jmath}} \ \equiv \ {1 \over \lm^{(\hat{\jmath})}(u)} \, {\partial \over \partial u^{\hat{\jmath}}} \, .
\label{nabvec}
\ee

A straightforward computation from (\ref{Dirac-eq-curvilinear}) leads to the covariant Dirac equation in the form
\be
\lt[\gm^{\hat{\mu}} \lt(\Pvec_{\hat{\mu}} - \hb \, \Gmvec_{\hat{\mu}}(U) \rt) - m\rt]\psi(U) & = & 0 \, ,
\label{Dirac-eq-curvilinear-1}
\ee
where
\be
\Pvec_{\hat{\mu}} & = & \pvec_{\hat{\mu}} + \Omvec_{\hat{\mu}}
\label{Pvec}
\ee
is the canonical momentum operator in curvilinear co-ordinates, with
\be
\pvec_{\hat{\mu}} & = & i \hbar \, \nabvec_{\hat{\mu}} \, ,
\label{pvec}
\nl
\Omvec_{\hat{\mu}} & = & i \hbar \lt[\nabvec_{\hat{\mu}} \ln \lt(\lm^{(\hat{1})}(u) \, \lm^{(\hat{2})}(u) \, \lm^{(\hat{3})}(u)\rt)^{1/2} \rt] \, ,
\label{Ovec}
\nl
i \, \Gmvec_{\hat{\mu}} & = & \bar{\Gmvec}^{(\rm S)}_{\hat{\mu}}
+ \gm^{\hat{l}} \, \gm^{\hat{m}} \, \bar{\Gmvec}^{(\rm T)}_{\ze[\hat{l}\hat{m}]} \, \dl^{\ze}{}_{\hat{\mu}} \, ,
\label{Spin-Connection}
\nl
\bar{\Gmvec}^{(\rm S)}_{\ze} & = & {1 \over 12} {}^F{}R^m{}_{jmk,0}(T) \, X^j \, X^k \, ,
\label{Spin-Connection-0-S}
\nl
\bar{\Gmvec}^{(\rm S)}_{\hat{\jmath}} & = & - \lt[{1 \over 2} {}^F{}R_{j00m}(T) + {1 \over 3} {}^F{}R_{jl0m,0}(T) \, X^l \rt] X^m \, ,
\qquad \hspace{1mm}
\label{Spin-Connection-j-S}
\nl
\bar{\Gmvec}^{(\rm T)}_{\ze[\hat{l}\hat{m}]} & = & 
{1 \over 2} \, {}^F{}R_{lm0k}(T) \, X^k \, . \qquad
\label{Spin-Connection-0-T}
\ee
For example, it is straightforward to verify in spherical co-ordinates $u^j = \lt(r, \th, \ph\rt)$ that \cite{Singh-Mobed2}
\be
\Pvec^{\hat{r}} & = & - \ihb \lt(\pt{r} + {1 \over r}\rt) \, , \quad
\Pvec^{\hat{\th}} \ = \ - {\ihb \over r} \lt(\pt{\th} + {1 \over 2} \, \cot \th\rt) \, ,
\nn
\Pvec^{\hat{\ph}} & = & - {\ihb \over r \, \sin \th} \, \pt{\ph} \, ,
\label{momentum-spherical}
\ee
where $\lm^{(\hat{1})}(u) = 1 \,$, $\lm^{(\hat{2})}(u) = r \,$, and $\lm^{(\hat{3})}(u) = r \, \sin \th \,$.

The final step follows by recalling the identity \cite{Aitchison}
\be
\gm^{\hat{\mu}} \, \gm^{\hat{\nu}} \, \gm^{\hat{\rho}} & = & \eta^{\hat{\nu} \hat{\rho}} \, \gm^{\hat{\mu}}
- 2 \, \gm^{[\hat{\nu}} \eta^{\hat{\rho}]\hat{\mu}} - i \, \gm^5 \, \gm^{\hat{\sg}} \, \varepsilon^{\hat{\mu} \hat{\nu} \hat{\rho}}{}_{\hat{\sg}} \, ,
\label{gm-identity}
\ee
where $\varepsilon^{\hat{\mu} \hat{\nu} \hat{\rho} \hat{\sg}}$ is the Levi-Civita symbol with
$\varepsilon^{\hat{0} \hat{1} \hat{2} \hat{3}} = 1$ \cite{Itzykson}.
Use of (\ref{gm-identity}) leads to a new expression for the spin connection, such that
\be
\Gmvec_{\hat{\mu}} & = &  \gm^5 \, \bar{\Gmvec}^{(\rm C)}_{\hat{\mu}} - i \, \bar{\Gmvec}^{(\rm S)}_{\hat{\mu}} \, ,
\label{Spin-connection-New}
\nl
\bar{\Gmvec}^{(\rm C)}_{\hat{\mu}} & = & \varepsilon^{\hat{0} \hat{l} \hat{m}}{}_{\hat{\mu}} \, \bar{\Gmvec}^{(\rm T)}_{\ze[\hat{l}\hat{m}]} \, ,
\label{Spin-Connection-C}
\ee
where the ``C'' in (\ref{Spin-Connection-C}) is the chiral-dependent part of the spin connection, while the ``S'' in
(\ref{Spin-Connection-0-S})--(\ref{Spin-Connection-j-S}) denotes the symmetric part under chiral symmetry.

\section{Spin-1/2 Particle Dirac Hamiltonian for General Motion in Curved Space-Time and {\em Zitterbewegung} Effects}
\label{sec:4}

Having obtained the required computations for the covariant Dirac equation in Fermi normal co-ordinates,
it is straightforward to determine from the
Schr\"{o}dinger equation $i \hbar \lt(\partial/\partial T \rt) \psi(U) = H \, \psi(U)$ that
\be
H & = &{} m \, \bt + \lt(\alvec \cdot \Pvec\rt) + \hb \lt[\sgvec^{\hat{\jmath}} \, \vec{\cal B}_j(T, \Xvec)
- {i \over 2} \, \alvec^{\hat{\jmath}} \, \vec{\cal E}_j(T, \Xvec) \rt]
\nn
&  &{} - {i \hb \over 3} \, {}^F{}R_{jlm0,0}(T) \, \alvec^{\hat{\jmath}} \, X^l \, X^m - {i \hb \over 12} \, {}^F{}R^m{}_{jmk,0}(T) \, X^j \, X^k
\label{Dirac-Hamiltonian}
\ee
is the Dirac Hamiltonian expressed in terms of the ``electric'' and ``magnetic'' field components \cite{Mashhoon3}
\be
\vec{\cal E}_j(T, \Xvec) & = & - {}^F{}R_{0j0k}(T) \, X^k \, ,
\label{E-Riemann}
\nl
\vec{\cal B}_j(T, \Xvec) & = & {1 \over 2} \, \varepsilon^{0lm}{}_j \, {}^F{}R_{lm0k}(T) \, X^k \, ,
\label{B-Riemann}
\ee
where $\partial_j \, \vec{\cal B}^j(T, \Xvec) = 0$ and $\partial_j \, \vec{\cal E}^j(T, \Xvec) = -{}^F{}R_{00}(T)$,
the analog of Poisson's equation in electromagnetism, whose matter field source generates
the Ricci curvature tensor in the Fermi frame.
The Dirac Hamiltonian (\ref{Dirac-Hamiltonian}) is comprised of operators which are defined in the local Lorentz frame
and are related to corresponding operators in general curvilinear co-ordinates in the standard way via the vierbein projection
functions (\ref{tetrad-coord-trans}) and (\ref{inv-tetrad-coord-trans}).
As such, the mathematical properties of these operators locally correspond precisely to the description given for
their counterparts in strictly flat space-time, with the understanding that the range of curvature deviation
satisfies the condition that $\lt|^F{}g_{\mu \nu}(X) - \eta_{\mu \nu}\rt| \ll 1$, particularly when
$X^j \rightarrow \Xvec_{(B)}^{\hat{\jmath}}(T)$.
Apart from the spin-dependent term coupled to (\ref{B-Riemann}), it is evident that the gravitational contributions to
(\ref{Dirac-Hamiltonian}) in its present form are anti-Hermitian.
If the time variations of the gravitational field are deemed small, then the last two terms in
(\ref{Dirac-Hamiltonian}) can be neglected.
However, for this paper all terms in the Dirac Hamiltonian are retained.

At this point, it is important to note that the Dirac Hamiltonian $H \rightarrow H_0 = m \, \bt + \lt(\alvec \cdot \Pvec\rt)$
as $X^j \rightarrow 0$, satisfying the weak equivalence principle for purely {\em classical} conceptions of space-time.
However, it so happens that if $X^j \rightarrow \Xvec_{(B)}^{\hat{\jmath}}(T)$ in the Fermi frame,
then a {\em quantum violation} of the weak equivalence principle emerges due to {\em Zitterbewegung}.
This can be shown explicitly by the following process, where for the sake of notational convenience the local Lorentz frame
indices will be unhatted for all subsequent computations.
Suppose the Dirac Hamiltonian is expressed as
\be
H & = & H_0 + H_{\rm G} \, ,
\label{H=H0+HG}
\ee
where $H_0$ is given by (\ref{H0}) and $H_{\rm G}$ is the interaction Hamiltonian involving space-time curvature.
Given that $\lt|X^j\rt| \sim \hbar/m$ and the curvature tensors in the Fermi frame are defined on the classical worldline
to satisfy $\lt|^F{}g_{\mu \nu}(X) - \eta_{\mu \nu}\rt| \ll 1$, it is reasonable to surmise that $H_{\rm G}  \ll H_0$.
If $X^j \rightarrow \Xvec_{(B)}^j(T)$  in the Fermi frame as {\em operators}, then the $X^l \, X^m$ within (\ref{Dirac-Hamiltonian})
become
\be
\fl
\Xvec_{(B)}^l(T) \, \Xvec_{(B)}^m(T) & = & {1 \over 2} \, \lt\{\Xvec_{(B)}^l(T) \, , \Xvec_{(B)}^m(T)\rt\}
+ {1 \over 2} \, \lt[\Xvec_{(B)}^l(T) \, , \Xvec_{(B)}^m(T)\rt] \, ,
\label{X^2-Barut}
\ee
the sum of anticommutator and commutator expressions, respectively, involving $\Xvec_{(B)}^l(T)$ and $\Xvec_{(B)}^m(T)$.
This leads to
\be
H_{\rm G} & = & {\hb \over 2} \, \varepsilon^{0lm}{}_j \, {}^F{}R_{lm0k}(T) \, \sgvec_{(B)}^j(T) \, \Xvec_{(B)}^k(T)
- {i \hb \over 2} \, {}^F{}R_{0j0k}(T) \, \alvec_{(B)}^j(T) \, \Xvec_{(B)}^k(T)
\nn
&  &{} - {i \hb \over 6} \, {}^F{}R_{j(lm)0,0}(T) \, \alvec_{(B)}^j(T) \, \lt\{\Xvec_{(B)}^l(T) \, , \Xvec_{(B)}^m(T)\rt\}
\nn
&  &{} + {i \hb \over 12} \, {}^F{}R_{j0lm,0}(T) \, \alvec_{(B)}^j(T) \, \lt[\Xvec_{(B)}^l(T) \, , \Xvec_{(B)}^m(T)\rt]
\nn
&  &{} - {i \hb \over 24} \, {}^F{}R^j{}_{ljm,0}(T) \, \lt\{\Xvec_{(B)}^l(T) \, , \Xvec_{(B)}^m(T)\rt\} \,
\label{HG}
\ee
from substitution of (\ref{X^2-Barut}) into the interaction Hamiltonian,
with ${}^F{}R_{j(lm)0}(T) = {1 \over 2} \lt[{}^F{}R_{jlm0}(T) + {}^F{}R_{jml0}(T)\rt]$
denoting symmetrization of the middle two indices, and using the cyclic permutation property of
indices for the Riemann tensor in the fourth term of (\ref{HG}).
Upon evaluation in terms of
\be
\lt\{\Xvec_{(B)}^l(T) \, , \Xvec_{(B)}^m(T)\rt\} & = & -{8 \over \omz^2} \, \sin^2 \lt(\omz T/2\rt) \, \eta^{lm} \, ,
\label{anticomm}
\nl
\lt[\Xvec_{(B)}^l(T) \, , \Xvec_{(B)}^m(T)\rt] & = & -{8 i \over \omz^2} \, \sin^2 \lt(\omz T/2\rt) \, \varepsilon^{0lm}{}_n \, \sgvec_0^n \, ,
\label{comm}
\ee
and with substitutions of (\ref{rel-X-Barut}), (\ref{alpha-Barut}), and (\ref{sigma-Barut}) into (\ref{HG}),
it follows that
\be
H_{\rm G} & = & -{\hb \over \omz} \, \sin^2 \lt(\omz T/2\rt) \lt\{ {}^F{}R_{00}(T) \rt.
\nn
&  &{} + \lt. 2 \lt[{}^F{}R_{0j}(T) + {4 \over 3 \, \omz} \, \sin\lt(\omz T\rt){}^F{}R_{0j,0}(T)\rt] \alvec_0^j \rt\} \bt_0
\nn
&  &{} + {i \hb \over \omz} \lt\{\sin \lt(\omz T\rt) \, {}^F{}R_{00}(T)
+ {1 \over 3 \, \omz} \, \sin^2 \lt(\omz T/2\rt) \, {}^F{}R^{jk}{}_{jk,0}(T) \rt.
\nn
&  &{}
- \lt[\sin\lt(\omz T\rt){}^F{}R_{0j}(T) \rt.
\nn
&  &{} + \lt. \lt. {8 \over 3 \, \omz} \, \sin^2 \lt(\omz T/2\rt) \cos \lt(\omz T\rt)
{}^F{}R_{0j,0}(T)\rt]\alvec_0^j \rt\} \, ,
\label{HG-1}
\ee
where the quantum violation of the weak equivalence principle formally appears due to the explicit coupling of
space-time curvature to the spin-1/2 particle's rest mass via the {\em Zitterbewegung} frequency $\omz = 2m/\hb$.
This is by no means the first time that a suggestion of weak equivalence violation has appeared in the literature.
For example, it was suggested years ago by Greenberger that, for a bound-state quantum mechanical particle
in an external gravitational potential, various violations of the weak equivalence principle
emerge due to the direct coupling of particle mass to the gravitational field that subsequently do not appear
in the classical limit \cite{Greenberger}.


Clearly, (\ref{HG-1}) vanishes as $\hb \rightarrow 0$ or $m \rightarrow \infty$,
while the Hermitian part of $H_{\rm G} \rightarrow 0$ and the anti-Hermitian part of $H_{\rm G}$ is regular as $m \rightarrow 0$.
In fact, it is shown that
\be
\lim_{m \rightarrow 0} H_{\rm G} & = & i \hb \lt\{\lt[{}^F{}R_{00}(T)
+ {1 \over 12} \, {}^F{}R^{jk}{}_{jk,0}(T) \, T \rt] \rt.
\nn
&  &{}
- \lt. \lt[{}^F{}R_{0j}(T) + {2 \over 3} \, {}^F{}R_{0j,0}(T) \, T\rt]\alvec_0^j \rt\} T \, ,
\label{HG-m=0-limit}
\ee
which is relevant for (almost) massless neutrinos.
It is particularly interesting to note that the first two terms in (\ref{HG-1}) coupled to $\bt_0$
form a gravitationally-induced effective mass, where ${}^F{}R_{00}(T)$ is likened to a potential energy term,
while the remaining terms serve as energy flux contributions with instantaneous velocity
$v_{\rm inst.} \equiv \lt|\alvec_0\rt| = 1$.
A similar identification exists for the anti-Hermitian terms not coupled to $\bt_0$,
which yield decay width contributions due to {\em Zitterbewegung} in curved space-time.

It is worthwhile to make a general comment about the anti-Hermitian contributions to (\ref{HG-1}).
Since it is practically true that the time scale associated with the Fermi frame curvature tensor
will be very much longer than the time scale corresponding to the {\em Zitterbewegung} frequency,
it is possible to regard $^F{}R_{\mu \nu \al \bt}(T)$ as effectively {\em constant} in time over a complete cycle
defined by $\omz$.
When this is taken into account, (\ref{HG-1}) reduces to
\be
H_{\rm G} & \rightarrow & -{\hb \over \omz} \, \sin^2 \lt(\omz T/2\rt) \lt\{ {}^F{}R_{00}(T) \rt.
\nn
&  &{} + \lt. 2 \lt[{}^F{}R_{0j}(T) + {4 \over 3 \, \omz} \, \sin\lt(\omz T\rt){}^F{}R_{0j,0}(T)\rt] \alvec_0^j \rt\} \bt_0
\nn
&  &{} + {i \hb \over 3 \, \omz^2} \, \sin^2 \lt(\omz T/2\rt) \lt[{}^F{}R^{jk}{}_{jk,0}(T) \rt.
\nn
&  &{} - \lt. 8 \, \cos \lt(\omz T\rt) {}^F{}R_{0j,0}(T) \, \alvec_0^j \rt] \, .
\label{HG-2}
\ee
If it so happens that the curvature tensors time-evolve adiabatically \cite{Parker} and
$\alvec_0^j \, \bt_0 \rightarrow {1 \over 2} \lt\{\alvec_0^j \, , \bt_0\rt\} = 0$, then the sole contribution to $H_{\rm G}$
is Hermitian.
Therefore, in a practical sense the anti-Hermitian terms in (\ref{HG-1}) are of little consequence in determining the
dynamics of a massive spin-1/2 particle in curved space-time due to {\em Zitterbewegung}.

\section{Time Evolution of Quantum Operators}
\label{sec:5}

Having now determined the Dirac Hamiltonian (\ref{H=H0+HG}) with {\em Zitterbewegung} contributions according to
(\ref{H0}) and (\ref{HG-1}), it is possible to compute the time evolution of quantum operators via the
Heisenberg equations of motion.
This can be done in perturbative form by solving for $\Avec \approx \Avec_{(0)} + \Avec_{(1)}$, where
\be
{\d \Avec_{(0)}(T) \over \d T} & \approx & {i \over \hb} \lt[H_0 \, , \Avec(T)\rt] \, ,
\label{dA-dt0}
\nl
\nn
{\d \Avec_{(1)}(T) \over \d T} & \approx & {i \over \hb} \lt[H_{\rm G} \, , \Avec(T)\rt] \, .
\label{dA-dt1}
\ee
%
%

By this approach, explicit computations for the time evolution of the relative position and momentum operators
can be determined in a gravitational background, where the specific case of quasi-circular motion
around a spherical black hole described by the Vaidya metric is considered in detail \cite{Singh}.
This eventually leads to additional contributions to the position-momentum measurement uncertainties
obtained in (\ref{Heisenberg-Barut}) due to space-time curvature and {\em Zitterbewegung}.
A separate computation for the time evolution of the canonical momentum $\Pvec^j(T) \approx \Pvec_{(0)}^j(T) + \Pvec_{(1)}^j(T)$
can also be determined, noting that in curvilinear co-ordinates the zeroth-order terms $\Pvec_{(0)}^j(T)$
are {\em not} constants of the motion.
When coupled with the periodic nature of the resulting vector differential equation, it follows that
solutions to $\Pvec^j(T)$ take the form given by Floquet's theorem \cite{Chicone1,Bransden}.
Finally, a formal presentation of the time derivative for the Pauli-Lubanski vector, the operator that describes
particle spin in covariant form, is given with respect to the perturbation Hamiltonian $H_{\rm G}$ and briefly discussed.

\subsection{Time Evolution of the Relative Position and Momentum in Curved Space-Time due to {\em Zitterbewegung}}

\subsubsection{Relative Position $\Xvec^j(T)$}

In the local rest frame, the relative position operator is
\be
\Xvec^j(T) & \approx & \Xvec_{(B)}^j(T) + \Xvec_{(1)}^j(T) \, ,
\label{X-rest-frame}
\ee
where $\Xvec_{(B)}^j(T)$ is given by (\ref{rel-X-Barut}) and
\be
\fl
{\d \Xvec_{(1)}^j(T) \over \d T} & = & {4 \over \omz^2} \, \sin^2 \lt(\omz T/2\rt)
\nn
&  &{} \times \lt\{ {}^F{}R_{00}(T) \lt[\sin^2 \lt(\omz T/2\rt) \alvec_0^j + {1 \over 2} \, \sin\lt(\omz T\rt) \lt(i \alvec_0^j \, \bt_0\rt)\rt] \rt.
\nn
&  &{} -i \lt[{8 \over 3 \, \omz} \, {}^F{}R^{0j}{}_{,0}(T) \, \sin^2 \lt(\omz T/2\rt) \bt_0 \rt.
\nn
&  &{}- \lt. \lt.
2 \lt[{}^F{}R_{0k}(T) + {2 \over 3 \, \omz} \, {}^F{}R_{0k,0}(T) \, \sin\lt(\omz T\rt) \rt] \varepsilon^{0jk}{}_l \, \sgvec_0^l \rt] \rt\}
\, .
\label{dX-dT1-1}
\ee
Retaining only the Hermitian part of (\ref{dX-dT1-1}) and integrating then leads to
\be
\Xvec_{(1)}^j (T) & = & {4 \over \omz^2} \int_0^T {}^F{}R_{00}(T') \, \sin^2 \lt(\omz T'/2\rt)
\nn
&  &{} \times \lt[\sin^2 \lt(\omz T'/2\rt) \alvec_0^j + {1 \over 2} \, \sin\lt(\omz T'\rt) \lt(i \alvec_0^j \, \bt_0\rt)\rt] \d T'
\, .
\label{X1}
\ee
Diagonalization of (\ref{X1}) then results in
\be
\lt. \Xvec_{(1)}^j (T) \rt|_{\rm diag.} & = & {4 \over \omz^2} \, \sgvec_0^j
\int_0^T {}^F{}R_{00}(T') \, \sin^3 \lt(\omz T'/2\rt) \d T' \, ,
\label{X1-diag}
\ee
where the squared magnitude of $\Xvec_{(1)}^j(T)$ before time averaging yields
\be
\lt|\Xvec_{(1)}(T)\rt|^2 & = & - \eta_{ij} \, \Xvec_{(1)}^i(T) \, \Xvec_{(1)}^j(T)
\nn
& = &  \lt| {4 \sqrt{3} \over \omz^2} \int_0^T {}^F{}R_{00}(T') \, \sin^3 \lt(\omz T'/2\rt) \d T' \rt|^2 \, .
\label{X1-magnitude-sq}
\ee

\subsubsection{Relative Momentum $\vec{\cal P}^j(T)$}

With the relative momentum described by
\be
\vec{\cal P}^j(T) & = & m \, \alvec^j(T) \ \approx \ \vec{\cal P}_{(B)}^j(T) + \vec{\cal P}_{(1)}^j(T) \, ,
\label{rel-P}
\ee
%
%
the time derivative of the first-order perturbation is
\be
\fl
{\d \vec{\cal P}_{(1)}^j(T) \over \d T} & = &
- {2 m \over \omz} \, {\cal F}(T) \lt[\sin\lt(\omz T\rt) \alvec_0^j + i \cos\lt(\omz T\rt) \alvec_0^j \, \bt_0\rt]
\nn
&  &{} + {2 i m \over \omz} \lt[\cos\lt(\omz T\rt) \vec{\cal C}^j(T) - \sin\lt(\omz T\rt) \vec{\cal D}^j(T)\rt]\bt_0
\nn
&  &{} - {2 i m \over \omz} \, \varepsilon^{0j}{}_{kl} \lt[\sin\lt(\omz T\rt) \vec{\cal C}^k(T) + \cos\lt(\omz T\rt) \vec{\cal D}^k(T)\rt]
\sgvec_0^l \, ,
\label{rel-dP-dt1}
\ee
where
\be
\fl
{\cal F}(T) & = & -\sin^2 \lt(\omz T/2\rt) \, {}^F{}R_{00}(T) \, ,
\label{F}
\nl
\fl
\vec{\cal C}_j(T) & = & -2 \, \sin^2 \lt(\omz T/2\rt) \lt[{}^F{}R_{0j}(T) + {4 \over 3 \, \omz} \, \sin\lt(\omz T\rt){}^F{}R_{0j,0}(T)\rt] \, ,
\label{Cj}
\nl
\fl
\vec{\cal D}_j(T) & = & - \lt[\sin\lt(\omz T\rt){}^F{}R_{0j}(T) \rt.
\nn
&  &{} + \lt. {8 \over 3 \, \omz} \, \sin^2 \lt(\omz T/2\rt) \cos \lt(\omz T\rt) {}^F{}R_{0j,0}(T)\rt] \, .
\label{Dj}
\ee
By also retaining only the Hermitian part of (\ref{rel-dP-dt1}) and integrating, it is shown that
\be
\fl
\vec{\cal P}_{(1)}^j(T) & = & - {2 m \over \omz}
\int_0^T {\cal F}(T') \lt[\sin\lt(\omz T'\rt) \alvec_0^j + \cos\lt(\omz T'\rt)\lt(i \alvec_0^j \, \bt_0\rt)\rt] \d T' \, ,
\label{rel-P1}
\ee
where in diagonalized form (\ref{rel-P1}) becomes
\be
\lt. \vec{\cal P}_{(1)}^j(T) \rt|_{\rm diag.} & = & {2 m \over \omz} \, \sgvec_0^j \int_0^T {}^F{}R_{00}(T')
\, \sin^2\lt(\omz T'/2\rt) \, \d T' \, .
\label{rel-P1-diag}
\ee
A corresponding computation for the squared magnitude of $\vec{\cal P}_{(1)}^j(T)$ prior to time averaging then results in
\be
\lt|\vec{\cal P}_{(1)}(T)\rt|^2 & = & - \eta_{ij} \, \vec{\cal P}_{(1)}^i(T) \, \vec{\cal P}_{(1)}^j(T)
\nn
& = &  \lt|{2 \sqrt{3} \, m \over \omz} \int_0^T {}^F{}R_{00}(T') \, \sin^2 \lt(\omz T'/2\rt) \d T' \rt|^2 \, .
\label{rel-P1-magnitude-sq}
\ee

\subsubsection{Position-Momentum Measurement Uncertainties in the Vaidya Background}

To illustrate the contribution of {\em Zitterbewegung} in curved space-time to the position-momentum
uncertainties in measurement, consider the quasi-circular motion of a spin-1/2 particle around
a spherically symmetric black hole in the presence of radiation, as described by the Vaidya metric \cite{Singh}.
For this computation, it is assumed that the background source $M$ monotonically changes in time according to
\be
M & = & M_0 + \Dl M(T) \, ,
\label{M}
\ee
where $M_0$ is the (static) Schwarzschild mass,
\be
\Dl M(T) & = & {\al \over A} \, \lt|\d \lt(\Dl M\rt) \over \d \xi \rt| \, T \, ,
\label{dM}
\nl
A^2 & = & 1 - {2 M_0 \over r} \, ,
\label{A-sq}
\ee
$r$ is the particle's orbital radius, and $\lt|\d \lt(\Dl M\rt)/\d \xi \rt|$
is the source's rate of change along $\xi$, a radial null co-ordinate.
If $\al = +1$, then $\xi$ is the {\em advanced null co-ordinate} corresponding to infalling radiation,
while $\al = -1$ implies that $\xi$ is the {\em retarded null co-ordinate} for outgoing radiation.
Then it is determined that \cite{Singh}
\be
{}^F{}R_{00}(T) & \approx & {2 \, \al \over N^2 \, r^2} \lt|\d \lt(\Dl M\rt) \over \d \xi \rt|
\lt(1 + 2 \, \al \lt|\d \lt(\Dl M\rt) \over \d \xi \rt| \rt.
\nn
&  &{} \times \lt. \lt[{\lt(r \OmK\rt) \over 2 \, N^6} \, C\lt(r, \OmK T\rt) - {2 \over N^2 A^2} \, \sin^2 \lt(\OmK T\rt) \rt] \rt) \, ,
\label{Ricci-Vaidya}
\ee
where $\OmK = \sqrt{M_0/r^3}$ is the Keplerian frequency of the orbit,
\be
N^2 & = & 1 - {3 M_0 \over r} \, , \qquad \lt(r \ > \ 3 M_0\rt)
\label{N-sq}
\ee
and
\be
C\lt(r, \OmK T\rt) & = & 2 \, \sin\lt(2 \, \OmK T\rt)
\nn
&  &{} + {N \over r \, \OmK} \lt[\lt(1 - 2 \, r \OmK\rt) \sin \lt(2 \, \OmK T\rt) - 2 \, \OmK T \rt] \, .
\label{C}
\ee
To leading order in $\lt|\d \lt(\Dl M\rt)/\d \xi\rt|$, it is straightforward to show that
\be
\fl
\lt\langle \Xvec_{(1)}^j(T) \rt\rangle_{\rm Vaidya} & \approx & {16 \over 3} \,
{\al \over N^2 \, r^2 \, \omz^2} \, \lt|\d \lt(\Dl M\rt) \over \d \xi \rt| \, \lt(\hb \over m\rt) \, \sgvec_0^j \, ,
\label{X1-magnitude-Vaidya-component}
\nl
\nn
\fl
\lt\langle \lt|\Xvec(T)\rt|^2 \rt\rangle_{\rm Vaidya} & \approx &
\lt\langle \lt|\Xvec_{(B)}(T)\rt|^2 \rt\rangle
\nn
&  &{} + {16 \, \al \over N^2 \, r^2 \, \omz^2} \, \lt|\d \lt(\Dl M\rt) \over \d \xi \rt| \, \lt(\hb \over m\rt)^2
\lt(- {8 \over 5 \pi} \, \gm_0^5 + i \, \bt_0 \, \gm_0^5\rt) \, ,
\label{X-magnitude-sq-Vaidya-time-avg}
\nl
\nn
\fl
\lt|\lt\langle \Xvec(T) \rt\rangle\rt|^2_{\rm Vaidya} & \approx & \lt|\lt\langle \Xvec_{(B)}(T) \rt\rangle\rt|^2
+ {16 \, \al \over N^2 \, r^2 \, \omz^2} \, \lt|\d \lt(\Dl M\rt) \over \d \xi \rt| \, \lt(\hb \over m\rt)^2
\lt(i \, \bt_0 \, \gm_0^5\rt) \, ,
\label{X-magnitude-Vaidya-time-avg-sq}
\ee
which results in
\be
\lt(\Dl \Xvec\rt)_{\rm Vaidya} & \approx & \lt(1 - {256 \over 15 \pi} \, {\al \over N^2 \, r^2 \, \omz^2}
\, \lt|\d \lt(\Dl M\rt) \over \d \xi \rt| \, \gm_0^5 \rt) \lt(\Dl \Xvec_{(B)}\rt) \,
\label{dX-Vaidya}
\ee
for the uncertainty in local position measurement.
Similarly, it becomes evident that
\be
\fl
\lt\langle \lt| \vec{\cal P}(T)\rt|^2 \rt\rangle_{\rm Vaidya} & \approx &
\lt\langle \lt| \vec{\cal P}_{(B)}(T)\rt|^2 \rt\rangle
- {18 \, \al \over N^2 \, r^2 \, \omz^2} \, \lt|\d \lt(\Dl M\rt) \over \d \xi \rt| \, \lt(i \, \bt_0 \, \gm_0^5\rt) m^2 \, ,
\label{rel-P-magnitude-sq-Vaidya-time-avg}
\nl
\nn
\fl
\lt|\lt\langle \vec{\cal P}(T) \rt\rangle\rt|^2_{\rm Vaidya} & = &
\lt|\lt\langle \vec{\cal P}_{(1)}(T) \rt\rangle\rt|^2_{\rm Vaidya}
\ \approx \ \lt({2 \, \sqrt{3} \, \pi \over N^2 \, r^2 \, \omz^2} \, \lt|\d \lt(\Dl M\rt) \over \d \xi \rt| \, m\rt)^2
\, ,
\label{rel-P-magnitude-Vaidya-time-avg-sq}
\ee
leading to
\be
\lt(\Dl \vec{\cal P}\rt)_{\rm Vaidya} & \approx & \lt(1 - {3 \, \al \over N^2 \, r^2 \, \omz^2}
\, \lt|\d \lt(\Dl M\rt) \over \d \xi \rt|\lt(i \, \bt_0 \, \gm_0^5\rt) \rt) \lt(\Dl \vec{\cal P}_{(B)}\rt) \,
\label{rel-dP-Vaidya}
\ee
for the corresponding uncertainty for relative momentum.

Therefore, the combination of (\ref{dX-Vaidya}) and (\ref{rel-dP-Vaidya}) leads to the gravitationally-modified
relative position-momentum uncertainty relation
\be
\fl
\lt(\Dl \Xvec\rt)\lt(\Dl \vec{\cal P}\rt)_{\rm Vaidya} & \approx &
\lt[1 - {\al \over N^2 \, r^2 \, \omz^2} \, \lt|\d \lt(\Dl M\rt) \over \d \xi \rt| \lt({256 \over 15 \pi} + 3 i \, \bt_0\rt)\gm_0^5  \rt]
\lt(\Dl \Xvec_{(B)}\rt) \lt(\Dl \vec{\cal P}_{(B)}\rt) \, ,
\nn
\label{Heisenberg-Vaidya}
\ee
whose diagonalized form, in terms of the chiral representation for the gamma matrices \cite{Itzykson}, is
\be
\fl
\lt. \lt(\Dl \Xvec\rt)\lt(\Dl \vec{\cal P}\rt)_{\rm Vaidya} \rt|_{\rm diag.} & \approx &
\lt[1 \mp \sqrt{9 + \lt(256 \over 15 \pi\rt)^2} \, {\al \over N^2 \, r^2 \, \omz^2} \, \lt|\d \lt(\Dl M\rt) \over \d \xi \rt|\rt]
\lt(\Dl \Xvec_{(B)}\rt) \lt(\Dl \vec{\cal P}_{(B)}\rt) \, ,
\nn
\label{Heisenberg-Vaidya-diag}
\ee
where the upper sign in (\ref{Heisenberg-Vaidya-diag}) refers to the right-handed spinor,
while the lower sign refers to the left-handed spinor.
It is interesting to note that when $\omz \rightarrow 0$, the absolute magnitude of
$\lt(\Dl \Xvec\rt)\lt(\Dl \vec{\cal P}\rt)_{\rm Vaidya}$ becomes infinitely large,
while $\omz \rightarrow \infty$ reduces (\ref{Heisenberg-Vaidya-diag}) to $ \lt(\Dl \Xvec_{(B)}\rt) \lt(\Dl \vec{\cal P}_{(B)}\rt)$.

Assuming the upper sign in (\ref{Heisenberg-Vaidya-diag}), the choice of $\al = +1$
shows that infalling radiation serves to reduce the uncertainty relation computed in flat space-time,
though $\lt|\d \lt(\Dl M\rt)/\d \xi\rt| \ll 1$ for realistic spherical mass accretion rates involving astrophysical
black holes, due to the Eddington luminosity limit \cite{Singh}.
As such, the gravitational contribution to the position-momentum uncertainty in measurement is negligibly
small for this circumstance.
However, it is unclear if this condition still applies where it concerns microscopic black holes,
especially if $r$ reduces to the Compton wavelength scale or smaller.

Finally, to ensure that the Heisenberg uncertainty relation is satisfied, the prefactor in front of
$\lt(\Dl \Xvec_{(B)}\rt) \lt(\Dl \vec{\cal P}_{(B)}\rt)$ in (\ref{Heisenberg-Vaidya-diag}) must remain nonzero.
Given (\ref{N-sq}), this condition implies that
\be
r & > & 3M_0 \lt[1 \pm \sqrt{9 + \lt(256 \over 15 \pi\rt)^2} \, {\al \over 9 M_0^2 \, \omz^2} \,
\lt|\d \lt(\Dl M\rt) \over \d \xi \rt|\rt] \, .
\label{r-min}
\ee
Again, assuming the upper sign and infalling radiation for (\ref{r-min}), it is interesting to observe
a growth in the lower bound for $r$ due to gravitational and {\em Zitterbewegung} effects.
Since $r = 3 M_0$ is the (unstable) photon orbit in Schwarzschild space-time, it is not surprising that
the lower bound will grow due to mass accretion from infalling radiation.
However, its explicit dependence on the {\em Zitterbewegung} frequency is unusual, a byproduct of
the quantum violation of the weak equivalence principle.

\subsection{Time Evolution of the Canonical Momentum due to {\em Zitterbewegung}}

When (\ref{dA-dt0}) and (\ref{dA-dt1}) are applied to the canonical momentum, the result is
%
%
%
\be
{\d \Pvec_j^{(0)}(T) \over \d T} & = & - \Om(T)_j{}^k \, \Pvec_k^{(0)}(T) \, ,
\label{dP-dt0}
\nl
\nn
{\d \Pvec_j^{(1)}(T) \over \d T} & = & - \Om(T)_j{}^k \, \Pvec_k^{(1)}(T) + \nabvec_j H_{\rm G}(T) \, ,
\label{dP-dt1}
\ee
where
\be
\Om(T)_j{}^k & = & \lt[\lt(\nabvec_l \ln \lm^{(k)}\rt) \dl_j{}^k
- \lt(\nabvec_j \ln \lm^{(k)}\rt) \dl^k{}_l \rt] \alvec_{(0)}^l(T) \, .
\label{Omega}
\ee
There are two very important points to note concerning (\ref{Omega}).
First, the fact that $\Om(T)_j{}^k \neq 0$ in curvilinear co-ordinates indicates that
the standard free-particle approach only applies for Cartesian co-ordinates, which implicitly
requires strictly inertial or rectilinear motion for the spin-1/2 particle.
For non-inertial motion with rotation, a more complicated time evolution applies.
The second point is that (\ref{Omega}) is {\em periodic} due to {\em Zitterbewegung} contributions from
$\alvec_{(0)}^j(T)$.
As a consequence, it follows that
\be
\Om(T + 2\pi/\omz)_j{}^k & = & \Om(T)_j{}^k \, ,
\label{Omega-periodic}
\ee
%
%
which satisfies the conditions for Floquet's theorem to apply \cite{Chicone1,Bransden}.
This leads to the free-particle solution of the form
\be
\Pvec_j^{(0)}(T) & \approx & e^{\lm \, \omz T} \, \Pivec_j^{(0)}(T) \, ,
\label{P0}
\ee
where
\be
\Pivec_j^{(0)}(T + 2\pi/\omz) & = & \Pivec_j^{(0)}(T) \, .
\label{Omega-periodic}
\ee
With explicit initial conditions, the free-particle solution for $\Pvec_j^{(0)}(T)$ is
\be
\Pvec_j^{(0)}(T) & \approx & \Pvec_j^{(0)}(0) + e^{\lm \, \omz T} \, \Pivec_j^{(0)}(T) - \Pivec_j^{(0)}(0) \, ,
\label{P0-intital}
\ee
where Re$(\lm) > 0$ in (\ref{P0-intital}) indicates an instability in the solution space for certain values of $T$
and/or regions of the curved space-time background where this condition may be satisfied.

To solve for $\Pvec_j^{(1)}(T)$ from (\ref{dP-dt1}), it is possible to multiply from the left by a time-dependent
invertible matrix $\mu(T)_j{}^k$, such that
\be
\mu(T)_j{}^k \lt(\nabvec_k H_{\rm G}(T)\rt) & = &
{\d \over \d T} \lt[\mu(T)_j{}^k \, \Pvec_k^{(1)}(T)\rt]
\nn
&  &{} - \lt[{\d \mu(T)_j{}^k \over \d T} - \mu(T)_j{}^l \, \Om(T)_l{}^k \rt] \Pvec_k^{(1)}(T) \, .
\label{integrating-factor}
\ee
Then so long as a solution exists for the matrix differential equation
\be
{\d \mu(T)_j{}^k \over \d T} - \mu(T)_j{}^l \, \Om(T)_l{}^k & = & 0 \, ,
\label{integrating-factor-condition}
\ee
which is possible to obtain using Floquet theory,
it becomes evident that $\mu(T)_j{}^k$ is an {\em integrating factor} for (\ref{integrating-factor}),
yielding the general solution
\be
\fl
\Pvec_j^{(1)}(T) & = & \lt[\mu(T)^{-1}\rt]_j{}^l \lt[\int_0^T \mu(T')_l{}^k \lt(\nabvec_k H_{\rm G}(T')\rt) \d T'
+ \mu(0)_l{}^k \, \Pvec_k^{(1)}(0)\rt] \,
\label{P1}
\ee
for the first-order perturbation due to the interaction Hamiltonian $H_{\rm G}$.

\subsection{Time Evolution of Covariant Spin due to {\em Zitterbewegung}}

The final set of computations concern the time evolution of the Pauli-Lubanski vector,
the operator describing covariant spin, involving {\em Zitterbewegung} in a curved space-time background.
For spin-1/2 particles, the Pauli-Lubanski vector in a local Lorentz frame is \cite{Singh-Mobed1,Singh-Mobed2}
\be
\Wvec^\mu(T) & = & -{1 \over 4} \, \varepsilon^\mu{}_{\al \bt \gm} \, \sg^{\al \bt}(T) \, \Pvec^\gm(T) \, ,
\label{W}
\ee
where $\sg^{\al \bt}(T)$ can be expressed as
\be
\sg^{\al \bt}(T) & = & 2i \, \dl^{[\al}{}_0 \, \dl^{\bt]}{}_j \, \alvec^j(T) - \varepsilon^{0\al\bt}{}_j \, \sgvec^j(T) \, .
\label{sg}
\ee
Applying (\ref{dA-dt0}) and (\ref{dA-dt1}) to (\ref{W}) 
eventually leads to
\be
\fl
{\d \Wvec_{(0)}^\mu(T) \over \d T} & = & -{m \over \hb} \, \varepsilon^\mu{}_{0ij} \, \alvec_{(0)}^i(T) \Pvec_{(0)}^j(T) \, \, \bt_{(0)}(T)
\nn
&  &{} + \eta^{\mu[i} \sgvec_{(0)}^{j]}(T) \lt(\nabvec_i \ln \lm^{(j)} \rt) \Pvec_j^{(0)}(T)
\nn
&  &{} + {i \over \hb} \, \dl^\mu{}_i \lt\{\lt[\Pvec_{(0)}^i(T) \, \Pvec^{(0)}_j(T) - \dl^i{}_j \lt(\Pvec^{(0)}_k(T) \, \Pvec_{(0)}^k(T)\rt)\rt]
\sgvec_{(0)}^j(T) \rt.
\nn
&  &{} + \lt. \hb \, \Rvec_{(0)}^i(T) \rt\} \, ,
\label{dW-dt0}
\nl
\fl
{\d \Wvec_{(1)}^\mu(T) \over \d T} & = & - {1 \over 4} \, \varepsilon^\mu{}_{\al \bt \gm} \, \sg^{\al \bt}(T) \lt(\nabvec^\gm H_{\rm G}(T)\rt)
\nn
&   & -{1 \over \omz} \, \varepsilon^\mu{}_{0j\gm}
\lt\{\cos\lt(\omz T\rt) \lt[{\cal F}(T) \, \alvec_0^j - \vec{\cal C}^j(T) \rt] \rt.
\nn
&  &{} + \lt. \sin \lt(\omz T\rt) \, \vec{\cal D}^j(T)\rt\} \Pvec_{(0)}^\gm(T) \, \bt_0
\nn
&  &{} + {4 \over \omz} \lt[\cos\lt(\omz T\rt) + \sin\lt(\omz T\rt)\rt] \dl^\mu{}_{[k} \, \eta_{l]\gm} \,
\vec{\cal C}^k(T) \, \sgvec_0^l \, \Pvec_{(0)}^\gm(T)
\nn
&  &{} + {i \over \omz} \, {\cal F}(T) \, \sin\lt(\omz T\rt) \, \varepsilon^\mu{}_{0j\gm} \, \alvec_0^j \, \Pvec_{(0)}^\gm(T)
\nn
&  &{} + {1 \over \omz} \lt(\eta^{\mu 0} \, \eta_{j \gm} - \dl^\mu{}_j \, \dl^0{}_\gm\rt)
\varepsilon^{0j}{}_{kl} \, \vec{\cal C}^k(T) \, \alvec_0^l \, \Pvec_{(0)}^\gm(T) \lt(i + \bt_0\rt) \,
\label{dW-dt1}
\ee
for the zeroth-order and first-order time derivatives of $\Wvec^\mu(T) \approx \Wvec_{(0)}^\mu(T) + \Wvec_{(1)}^\mu(T)$,
where
\be
\Rvec_{(0)}^j(T) & = &  {i \over 2\hbar} \, \varepsilon^{0jkl} \,
\lt[\Pvec^{(0)}_k(T), \Pvec^{(0)}_l(T)\rt]
\nn
& = & \varepsilon^{0jkl} \, \lt(\nabvec_k \, \ln \lm^{(l)}\rt) \Pvec^{(0)}_l(T) \,
\label{R}
\ee
in (\ref{dW-dt0}) is the {\em non-inertial dipole operator} \cite{Singh-Mobed1,Singh-Mobed2,Singh1},
which identically vanishes for momentum operators in Cartesian co-ordinates.
Because of the explicit and widespread dependence of $\Pvec_{(0)}^j(T)$ in both (\ref{dW-dt0}) and (\ref{dW-dt1}),
it is possible to foresee instabilities emerging in the time evolution of $\Wvec^\mu(T)$
due to instabilities in (\ref{P0-intital}).


\section{Conclusion}
\label{sec:6}

This paper explores the consequences of spin-1/2 particle {\em Zitterbewegung} in a
general curved space-time background using Fermi normal co-ordinates, where the spatial
vectors $X^j$ orthogonal to the classical worldline are treated as quantum fluctuations $\Xvec_{(B)}^j(T)$
in a local frame, according to the formalism first introduced by Barut and Bracken \cite{Barut2}.
An overarching conclusion is that the associated Dirac Hamiltonian $H_{\rm G}$ reveals a quantum
violation of the weak equivalence principle due to {\em Zitterbewegung}, confined to a worldtube
of radius $\omz^{-1}$.
Apart from one anti-Hermitian term in (\ref{HG-1}) that is coupled to ${}^F{}R^{jk}{}_{jk,0}(T)$, all the
gravitational contributions in $H_{\rm G}$ are dependent on the Ricci curvature tensor and its time derivative
in the Fermi frame.
In the classical limit as $\hb \rightarrow 0$, it follows that $H_{\rm G} \rightarrow 0$, as expected.
It should also be noted that, while the definition of quantum operators employed in this paper
is sufficient to demonstrate the possible existence of {\em Zitterbewegung} effects in a curved
space-time background, a more mathematically rigourous definition may be required to illustrate
the full extent of these effects.
This is a worthwhile consideration for future research on this topic.

When applied to the special case of quasi-circular orbital motion in the presence of a spherical
black hole in radiation, as described by the Vaidya metric, it is shown that uncertainties in the
simultaneous measurement of the particle's locally determined position and momentum are modified
due to the coupling of $\omz$ with the black hole's mass rate-of-change $\lt|\d \lt(\Dl M\rt)/\d \xi \rt|$,
with interesting physical consequences.
By further allowing a description of the problem in terms of general curvilinear co-ordinates
to accommodate non-inertial motion along the worldline, the purely flat space-time treatment
of this problem becomes generalized by showing that the canonical momentum is subject to the Floquet
theory of differential equations involving oscillation frequency $\omz$, with potential stability implications
due to {\em Zitterbewegung}.

There are numerous potentially significant consequences and future directions worth exploring
from the main expressions presented in this paper.
For example, the fact that $H_{\rm G}$ is explicitly time-dependent due to {\em Zitterbewegung}
indicates the possibility of identifying transition rates for local energy states about the
spin-1/2 particle's classical worldline \cite{Bransden}.
While the gravitational and {\em Zitterbewegung} contributions do not survive for a {\em classical}
vacuum space-time background, where ${}^F{}R_{\mu \nu}(T) = 0$ identically, this may not necessarily
be true if zero-point fluctuations and vacuum polarization effects \cite{Birrell}
due to the Casimir effect, for example, appear as a consequence of virtual pair-production.
Given that QED predicts the existence of vacuum bubbles that have no observable effect in purely flat space-time,
it is reasonable to consider whether this property holds true in the local presence of space-time curvature.
Finally, the results presented here may have potential cosmological implications in terms of the
very small value for the (observed) cosmological constant, with the perceived global acceleration of
the Universe as an observable effect.
These and other possibilities may be explored at a later date.

\section{Acknowledgements}

The authors wish to thank Bahram Mashhoon from the University of Missouri for helpful comments and suggestions.
They are also grateful to the Faculty of Science at the University of Regina for financial support.

\end{document}